%% file: minosrmulti.tex
\documentclass[%
 reprint,superscriptaddress,letterpaper,floatfix,nofootinbib,
showpacs,
preprintnumbers,
 amsmath,amssymb,
 aps,prd
]{revtex4-1}

\usepackage{graphicx}
\usepackage{dcolumn}
\usepackage{bm}
\usepackage[mathlines]{lineno}

\usepackage{varwidth}
\usepackage{xspace}
\usepackage{multirow}
\usepackage{rotating}
\usepackage{array}
\usepackage{color}
\usepackage{colortbl}
\usepackage{appendix}
\usepackage{setspace}
\usepackage[]{units}
\definecolor{Gray}{gray}{0.8}

\begin{document}

\preprint{FERMILAB-PUB-15-568-ND}

\title{Measurement of the multiple-muon charge ratio in the MINOS Far Detector}

\input{ab}

\date{\today}

\begin{abstract}
The charge ratio,  $R_\mu = N_{\mu^+}/N_{\mu^-}$, for cosmogenic 
multiple-muon events observed at an underground depth of 2070 mwe has been 
measured using the magnetized MINOS Far Detector.  The multiple-muon events, 
recorded nearly continuously from August 2003 until April 2012, comprise two 
independent data sets imaged with opposite magnetic field polarities, the comparison 
of which allows the systematic uncertainties of the measurement to be minimized.      
The multiple-muon charge ratio is determined to be
$R_\mu = 1.104 \pm 0.006 {\rm \,(stat.)} ^{+0.009}_{-0.010} {\rm \,(syst.)} $.
This measurement complements previous determinations of single-muon and 
multiple-muon charge ratios at underground sites and serves to constrain models of 
cosmic-ray interactions at TeV energies.

\end{abstract}


\maketitle

\section{\label{sec:intro} Introduction}

Atmospheric muons are produced when primary cosmic-ray nuclei interact in the upper 
atmosphere, yielding hadronic showers that contain pions and kaons. These secondary 
mesons can either interact in further collisions in the atmosphere or decay to produce 
atmospheric muons. Precision measurements of the muon charge ratio, 
$R_\mu \equiv N_{\mu^+}$/$N_{\mu^-}$, in cosmic-rays can be used to improve models 
of the interactions of cosmic rays in the atmosphere. Single-muon and multiple-muon 
events provide complementary information that feeds into the development of these 
models. In addition, measurements of the cosmic-ray muon charge ratio from a few GeV 
to a few TeV are important for constraining calculations of atmospheric neutrino fluxes.  
These are of interest both for detailed measurements of neutrino oscillations in 
atmospheric neutrino experiments and also for calculations of backgrounds for neutrino 
telescopes.  The muon charge ratio is a particularly useful tool for testing the predicted 
atmospheric $\nu/\bar{\nu}$ ratio.

Single-muon charge ratio measurements performed by 
MINOS (Near Detector) \cite{bib:adamson1},  L3+C \cite{bib:achard}, 
Bess-TeV \cite{bib:haino}, CosmoALEPH \cite{bib:zimmermann} and 
CMS \cite{bib:khachatryan} at surface-level energies, $E_\mu$, 
ranging from a few hundred MeV to \unit[100]{GeV} are consistent 
with the 2001 world average of $1.268 
\pm [0.008 + 0.0002 \: E_\mu/{\rm GeV}]$ \cite{bib:hebbeker}. This apparent constancy 
over three orders of magnitude in muon energy can be interpreted as a consequence of 
Feynman scaling \cite{bib:feynman}. At TeV surface energies, MINOS (Far 
Detector) \cite{bib:adamson2} and OPERA \cite{bib:opera2} reported higher values for 
the muon charge ratio, $1.374 \pm 0.004 {\rm \,(stat.)}^{+0.012}_{-0.010}{\rm\,(syst.)}$ 
and $1.377 \pm 0.006{\rm\,(stat.)}^{+0.007}_{-0.001}{\rm\,(syst.)}$, respectively. The 
atmospheric muon charge ratio for single muons is not unity because the primary cosmic 
rays are mostly protons, which have a preponderance of $u$ quarks, favoring the 
production of a leading $\pi^+$ or $K^+$ over $\pi^-$ and $K^-$. The existence of 
associated production, {\it e.g.}, $K^+\Lambda$, additionally favors $K^+$ over $K^-$.  
Due to the steeply falling primary cosmic-ray energy spectrum, which follows an 
$E^{-2.7}$ power law, a single-muon event in a deep underground detector is more likely 
to arise from the decay of a leading hadron than from a secondary hadron or later 
generation hadrons. The rise at TeV energies is explained in Ref.~\cite{bib:schreiner} as 
an increased contribution from kaon decay in the region of muon energy and zenith 
angle $ \epsilon_\pi < E_\mu \cos{\theta_z} < \epsilon_K$. The critical energies, 
$\epsilon$, are meson energies for which the decay probability and interaction 
probability are equal at the altitude in the atmosphere where the majority of detected 
muons are produced. The values for these energies are $\epsilon_\pi = 115$~GeV and 
$\epsilon_K = 850$~GeV \cite{bib:zatsepin}. 

In underground detectors, a multiple-muon event occurs when two or more 
almost-parallel muons are observed that originate from a common cosmic-ray primary. 
The process typically involves more than the decay of a single leading hadron. Events 
can be produced by two or more hadrons from the first interaction, or by particles produced 
in secondary interactions or deeper in the hadronic shower. Some events are also 
produced by the dimuon decay of a single leading hadron, but the branching fraction for 
this process is relatively small. In the MINOS Far Detector, which has a depth of 2070 
meters-of-water equivalent (mwe), multiple-muon events account for 7\% of the 
observed cosmic-ray events. In a multiple-muon event, there can be some muons for 
which the charge is well measured and other muons for which the charge measurement 
is ambiguous. This paper reports the charge ratio in MINOS for tracks in multiple-muon 
events in which at least one muon's charge is well-measured, whether or not the 
charges of other muons in the same multiple-muon event are known. In multiple-muon 
events, all muons with a well-measured charge are included in the calculation of the 
charge ratio.

Previously, OPERA reported values of 
$1.23 \pm 0.06{\rm\,(stat.)}^{+0.017}_{-0.015}{\rm\,(syst.)} $ (2010) \cite{bib:opera1} 
and $1.098 \pm 0.023{\rm\,(stat.)}^{+0.015}_{-0.013}{\rm\,(syst.)}$ 
(2014) \cite{bib:opera2} for the multiple-muon charge ratio at a depth of \unit[3800] mwe, 
smaller than the single-muon ratio cited above. In the next three paragraphs, three 
related factors are discussed that might make the measured multiple-muon charge ratio 
lower than the single-muon charge ratio: (a) the importance of the leading $u$ quark 
charge is diminished for nonleading hadrons and those produced after the first 
interaction, (b) the possibility of an increased heavy-nucleus component of the 
cosmic-ray flux at high energy, and (c) the kinematics of multiple-muon events coupled 
with the maximum detectable momentum (MDM) of a magnetic detector like MINOS.

The single-muon charge ratio is larger than unity because the incoming cosmic rays  
have more $u$ quarks than $d$ quarks. In the production of additional positive and 
negative hadrons in the first and subsequent interactions, that effect must be diminished. 

A second effect comes from the fact that heavier elements make a relatively larger 
contribution to the cosmic-ray primaries responsible for multiple-muon events than for 
single-muon events, for two reasons. First, the mean primary cosmic-ray energy for 
observed multiple muons is higher than that for single muons, and it is expected that 
heavier elements become a more important component of cosmic-ray primaries at 
higher energies~\cite{bib:pdg14}. Second, massive primaries generate more 
high-energy muons per event than proton primaries of the same total energy. This is 
because the first interaction point of the heavy primary is likely to be higher in the 
atmosphere than for a proton primary.  A heavy nucleus has a larger cross section for 
the interactions with air, and the lower density in the upper atmosphere favors pion 
decay over interaction early in the cascade development. Heavy nuclei also contain 
neutrons, which have twice as many $d$ quarks as $u$ quarks, and are therefore more 
likely to produce a leading negative pion, resulting in a decreased muon charge ratio.

The third effect arises since the probability of being able to measure the curvature 
sufficiently well decreases with increasing muon momentum. A magnetic detector can 
only reliably measure the charge of muons with a momentum below the MDM, which 
depends on the magnetic field and the detector geometry. Thus sometimes the 
highest-energy muon at the detector in a multiple-muon event will not have the sign of 
its curvature determined. In these situations, only lower-energy muons, from nonleading 
pions, will be used in the determination of the charge ratio. Since the leading pion is the 
most likely to carry the excess positive charge in the shower, this effect will reduce the 
measured charge ratio. This effect is explained in more detail in 
Sec. \ref{sec:minos_FD}.

The paper is organized as follows: the MINOS Far Detector is described in 
Sec. \ref{sec:minos_FD}.  The analyses of the MINOS multiple-muon data and the 
Monte Carlo (MC) simulation are described respectively in Secs. \ref{sec:data} and 
\ref{sec:mc}. The determination of the multiple-muon charge ratio is presented in 
Sec. \ref{sec:det_mcr}, including the corrections for charge misidentification and the 
calculations of systematic uncertainties.  A summary is given in 
Sec. \ref{sec:summary}.


\section{The MINOS Far Detector}
\label{sec:minos_FD}

The MINOS Far Detector (FD) is a magnetized planar steel-scintillator tracking 
calorimeter located at a depth of \unit[2070]{mwe} in the Soudan Underground 
Laboratory, in an iron mine in northern Minnesota (latitude 47.82027$^\circ$ N and 
longitude 92.24141$^\circ$ W). The detector consists of two supermodules separated by 
a gap of \unit[1.15]{m} and has a total dimension of $\unit[8.0 \times 8.0 \times 31]{m^3}$. 
The two supermodules contain a total of 486 octagonal steel planes, each \unit[2.54]{cm} 
thick, interleaved with 484 planes of \unit[1]{cm} thick extruded polystyrene scintillator 
strips, at a \unit[5.94]{cm} pitch. Each scintillator plane has 192 strips of width 
\unit[4.1]{cm}. The scintillator strips in alternating detector planes are oriented at 
$\pm$45$^\circ$ to the vertical.  Each plane has a small hole in the center for the 
magnet coil.

Scintillation light is collected by wavelength-shifting (WLS) plastic fibers embedded 
within the scintillator strips. The WLS fibers are coupled to clear optical fibers at 
both ends of a strip and are read out using 16-pixel multianode photomultiplier tubes 
(PMTs). The signals from eight strips, each one of which is separated by approximately 
\unit[1]{m} within the same plane, are optically summed and read out by a single PMT 
pixel. The fibers summed on each pixel are different for the two sides of the detector, 
which enables the resulting eightfold ambiguity to be resolved for single-track events. 
For multiple-muon events, ambiguities are resolved with a high level of accuracy
using additional information from timing and event topology.

The data acquisition and trigger have been described in Ref.~\cite{bib:adamson3}. 
Time and pulse height on each strip are digitized locally.  The primary trigger requires 
activity to be observed on 4 planes out of 5 within \unit[156]{ns}. More detailed detector
information can be found in Ref.~\cite{bib:michael}.

In order to measure the momentum of muons traversing the detector, the steel has 
been magnetized into a toroidal field configuration. The field varies in strength from 
\unit[1.8]{T} near the magnetic coil to around \unit[1]{T} near the edges. In one magnetic 
field setting, negative muons resulting from interactions of neutrinos from the Fermilab
NuMI beam are focused toward the center of the detector. This magnetic field orientation 
will be referred to as the forward field (FF) configuration. In the reverse field (RF) 
configuration, the coil current is reversed and positive muons from beam antineutrinos 
are focused into the detector. 

A reconstruction program turns scintillator hits into tracks, and a Kalman filter 
procedure \cite{bib:fruhwirth} is used to fit the track trajectories. The Kalman filter 
performs a series of recursive matrix manipulations to specify the trajectory of 
the particle as well as the ratio of its charge to its momentum, $q/p$. It also 
provides an uncertainty, $\sigma({q/p})$, on the measured value of $q/p$. Single muon
tracks are found with high efficiency. The reconstruction program has not been tuned for 
multiple-muon events. The techniques to achieve charge separation for reconstructed 
muons are described in the next section, and the efficiencies for track reconstruction 
and charge separation are considered in Sec. \ref{sec:mc}.

One important aspect of the present analysis is the detector's MDM. Due to the leading 
particle effect mentioned in Sec. \ref{sec:intro}, the excess of positive charge is most 
likely carried by the highest-energy muon, which for a multiple-muon event is frequently 
the least likely to have a well-measured charge.  In MINOS, the MDM is approximately 
a function only of the angle with respect to the detector axis and the distance of closest 
approach to the magnet coil, called the impact parameter, which can vary from zero to 
four meters \cite{bib:fields}. For favorable values of these two parameters the MDM 
reaches \unit[470]{GeV}, but is as low as \unit[15]{GeV} for other angles and impact 
parameters. The 3$\sigma$ requirement on the measurement of curvature  in this 
analysis leads to a charge measurement for only a small fraction of single and multiple 
muons in MINOS (see Ref.~\cite{bib:adamson2} and Table~\ref{tab:select}). For a track 
20$^\circ$ from the zenith, the MDM varies from \unit[220]{GeV} for a track with an 
impact parameter of \unit[0.5]{m} mostly perpendicular to the magnetic field, to 
\unit[17]{GeV} for a track with an impact parameter of \unit[3.5]{m} mostly parallel to the 
magnetic field.


\section{Data Sample}
\label{sec:data}

The multiple-muon sample reported in this paper was recorded between August 2003 
and April 2012. During the data-taking period, the detector ran 80.97\% in the FF and 
19.03\% in the RF configurations.

Selection criteria are chosen to ensure good quality data, filter well-reconstructed 
multiple-muon events, and separate muons based on their charge. An initial 
preselection stage of the event selection aims to identify and remove periods of data 
associated with detector hardware problems \cite{bib:blake}. Events with two or more 
tracks are then selected for analysis. Next, a series of six track analysis cuts are 
applied to the data. First, the collection of multi-GeV muons within a multiple-muon 
event must be highly parallel; to ensure this condition, at least two muons must be 
reconstructed with an angular separation of less than 5$^\circ$. If at least two tracks in 
an event satisfy this cut, all the muons in that event may be counted in the multiplicity, 
$M$. 

Tracks are required to have crossed at least 20 planes in the detector, and to have a 
path length of at least \unit[2]{m}. Each track in a multiple-muon event must be 
reconstructed as pointing downward, based on timing in the scintillator. The entry point 
of each track is required to be less than 50\,cm from the outside surface of the detector 
and greater than 50\,cm from the central axis (referred to as the fiducial volume cut). 
To ensure the quality of track reconstruction, a selection requirement of $\chi^2/ndf < 2$
is placed on the goodness of fit variable returned by the Kalman filter procedure.
These selection cuts are similar to those used in the previous ND and FD single-muon 
charge ratio analyses \cite{bib:adamson1,bib:adamson2}. The multiplicity of an event is 
defined as the number of tracks passing these cuts.

The method to identify tracks with well-determined charge is the same as that used in 
the MINOS single-muon charge-ratio analysis \cite{bib:adamson2}, and is described in 
the rest of this section. This charge-separation procedure only selects a small fraction 
of tracks since many muons in MINOS do not noticeably bend in the magnetic field. In 
this paper, the charge ratio is defined for all tracks that are determined to have 
well-measured charge. If more than one track in a multiple-muon event has a 
well-measured charge, each will be included in the calculation of $R_\mu$. 91.6 \% of 
the events in the full multiple-muon sample satisfy this criterion.

\begin{figure}[tb]
\centering
\includegraphics[width=0.48\textwidth]{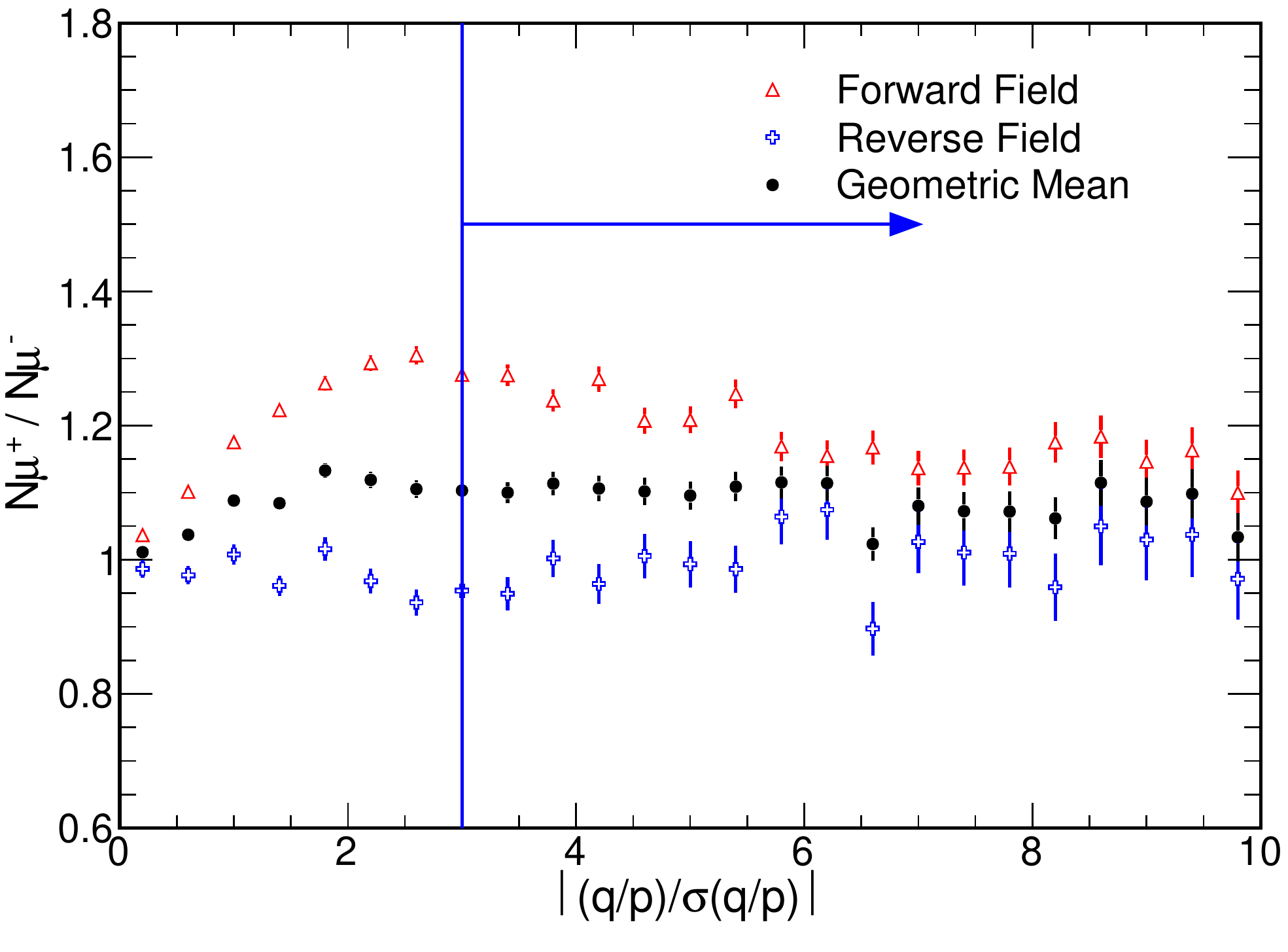}
\caption{Charge ratio for reconstructed multiple-muon tracks as a function of curvature 
significance after applying all other selection cuts. The vertical line denotes the 
minimum value for tracks used in the charge ratio measurement.}
\label{fig:resol}
\end{figure}

Two selection variables are used to increase the degree of confidence in the assigned 
curvature and charge sign of the tracks. The first variable uses outputs of the Kalman 
filter technique used in the track curvature fitting. The quantity $(q/p)/\sigma(q/p)$, 
called the curvature significance, can be thought of as the significance with which a 
straight-line fit to the track can be rejected, using the pattern of curvature that is 
expected given the magnetic field. Figure~\ref{fig:resol} shows the measured 
multiple-muon charge ratio in the data as a function of the curvature significance. The 
figure shows separately the data taken in the two magnetic field orientations, illustrating 
systematic differences in the charge ratio measurements between FF and RF data. 
These differences come from acceptance effects due to the magnetic field, detector
asymmetry, and detector alignment errors. To remove these biases, data taken in the 
two field orientations is combined by calculating a geometric mean (GM) between the 
two data sets, described at the beginning of Sec. \ref{sec:det_mcr}.

Events with low values of the curvature significance are typically high-momentum 
tracks ($>\unit[100]{GeV/c}$) that do not significantly curve while traversing the MINOS 
detector. For such tracks, the charge sign determined by the fitter becomes less reliable. 
As the curvature significance tends to zero, the fitter picks the two charge signs with 
nearly equal probability and, as can be seen from Fig.~\ref{fig:resol}, the measured 
charge ratio tends to unity. A cut is applied such that only tracks with 
$|(q/p)/\sigma(q/p)|>3$ are used in the analysis. 

The second charge quality selection variable, $BdL$, is defined as
\begin{equation}
BdL \equiv \int^{\rm end}_{\rm beg} | {\vec B}(r) \times {\vec n} | ~ dL,
\end{equation}
where $| {\vec B}(r) \times {\vec n} |$ is the component of the magnetic field 
perpendicular to the track direction, $ {\vec n}$, at a given point along the track path, 
$r$ is the distance from the detector center axis, $dL$ is the differential path length 
element along the track in the magnetic field, and the integral runs from the point where 
the muon enters the detector to the point at which it either exits the detector or stops in 
the detector. This $BdL$ variable quantifies the magnitude of the bending due to the 
magnetic field.

\begin{figure}[tb]
\centering
\includegraphics[width=0.48\textwidth]{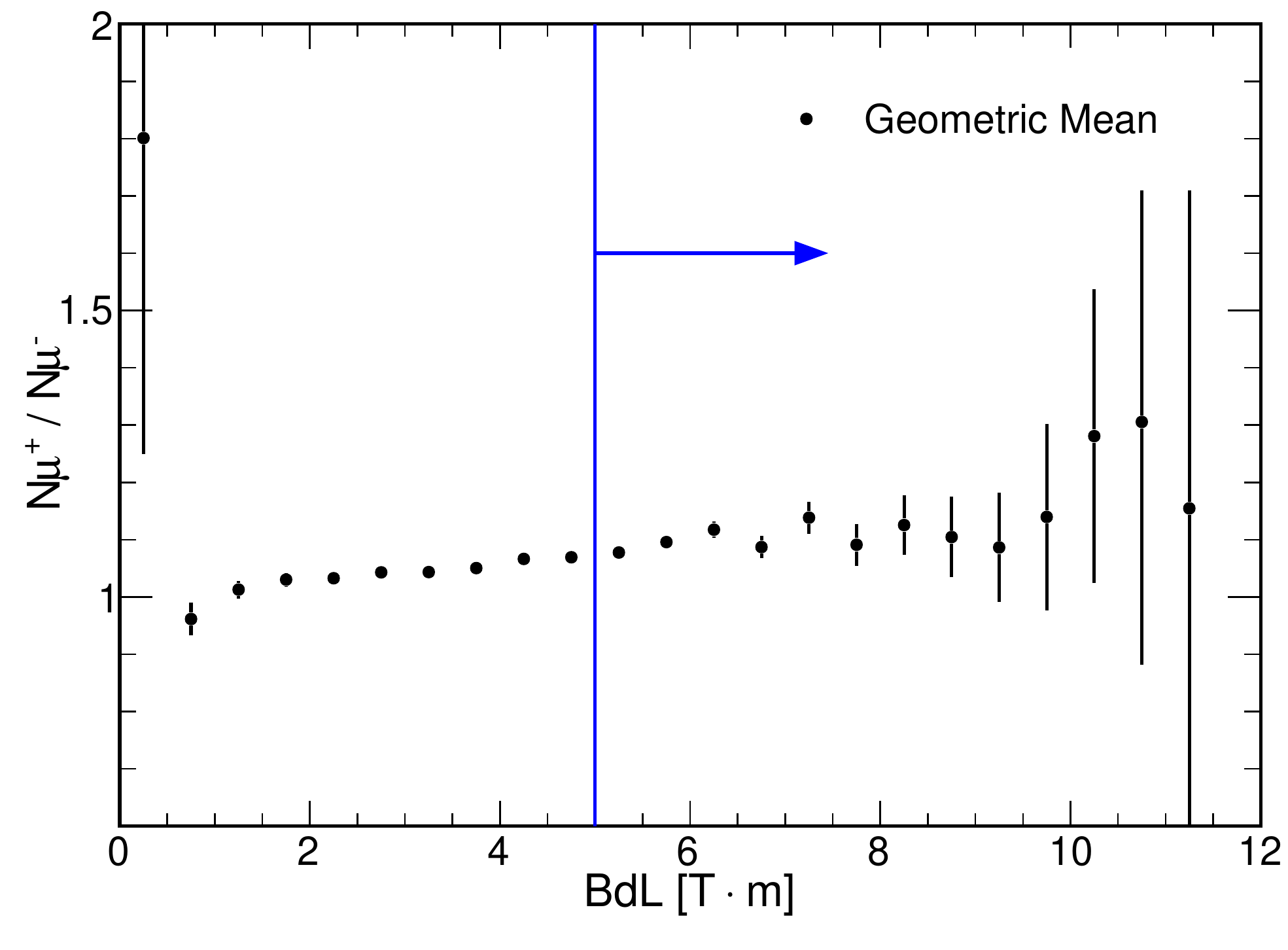}
\caption{Charge ratio as a function of $BdL$ for multiple-muon tracks passing the 
curvature significance cut. The vertical line shows  the minimum value for tracks used in 
the charge-ratio measurement.}
\label{fig:bdl}
\end{figure}

Figure~\ref{fig:bdl} shows the measured multiple-muon charge ratio in the data as a 
function of $BdL$. For this analysis, it was required that $BdL >$ \unit[5]{T$\cdot$m}. 
For low values of $BdL$, track curvature due to multiple scattering is comparable to the 
magnetic bending and the measured charge ratio approaches unity as expected in the 
case of random charge determination. The $BdL$ cut was chosen in Ref.
\cite{bib:adamson2} as the value above which charge misidentification becomes 
negligible.\footnote{In Ref.~\cite{bib:adamson2}, the length was defined as the total track 
length.  In Ref.~\cite{bib:adamson1} and in this paper, the length through the magnetized 
steel is used.  The cut was commensurately adjusted.} This issue is discussed in some 
detail in Ref.~\cite{bib:schreiner}.

Table~\ref{tab:select} summarizes the number of muon tracks that pass each of the 
selection cuts. The final selected sample is then used in the calculation of the 
multiple-muon charge ratio, which is described in Sec. \ref{sec:det_mcr}.

\begin{table}[h]
\caption{\label{tab:select} Summary of the applied cuts. Each row shows the total 
number of muons in both field configurations remaining after each successive cut is 
applied to the data. The numbers in parentheses show the percentage of muons 
remaining.}
\begin{center}
\small
\begin{tabular}{lcc}
\hline \hline
Cuts & \multicolumn{2}{c}{Number of muons remaining}\\
\hline
\\
Preselected tracks & \quad $8.35 \times 10^6$ & (100\%)\\
& & \\
Track analysis cuts & & \\
~~ Parallel tracks ($< 5^{\circ}$)  & \quad $7.31 \times 10^6$ & (87.5\%) \\ 
~~ 20 planes & \quad $5.88 \times 10^6$ & (70.5\%) \\
~~ 2 m track length  & \quad $5.87 \times 10^6$ & (70.3\%) \\
~~ Downward-going track & \quad $5.86 \times 10^6$ & (70.2\%) \\
~~ Fiducial volume & \quad $5.75 \times 10^6$ & (68.9\%) \\ 
~~ Fit quality: $\chi^2/ndf < 2$ & \quad $5.17 \times 10^6$ & (61.9\%) \\ 
& &\\
Charge-sign quality cuts & & \\
~~ $|(q/p)/\sigma(q/p)| > 3$ & \quad $1.08 \times 10^6$ & (13.0\%) \\
~~ $BdL >$ \unit[5]{T$\cdot$m} & \quad $3.12 \times 10^5$ & (3.7\%) \\
\hline \hline
\end{tabular}
\end{center}
\end{table} 


\section{Simulated atmospheric muons}
\label{sec:mc}

Two distinct samples of simulated atmospheric muons are used to evaluate the  
efficiency of the cuts in Table~\ref{tab:select}: one sample to evaluate the multiple-muon 
track reconstruction efficiency, and another sample to evaluate the charge separation 
efficiency. These two Monte Carlo (MC) data samples use different methods to simulate 
the muon charges and momenta, as well as the vertex positions where the muons enter 
the detector. In each case, this information is used as the input to the GEANT4-based 
simulation \cite{bib:geant} that propagates the muons through the MINOS FD, taking 
into account the magnetic field and the muon energy losses as they travel through the 
steel and scintillator. This GEANT4-based simulation furthermore models the production 
of light in the scintillator strips and the full chain of PMTs and readout electronics that 
converts this light into raw detector data. These simulated data are then passed through 
the same reconstruction and analysis as the real data.

\begin{table*}[!htb]
\caption{\label{tab:trk_eff} The number of reconstructed simulated multiple-muon 
events per multiplicity before charge separation. These tracks satisfy the analysis cuts in 
Table~\ref{tab:select}. The efficiency ${\varepsilon_{[rec.,\, {\it M}]}}$  is the fraction of 
events with a reconstructed multiplicity greater than 1. The efficiency 
${\varepsilon_{[rec.=gen]}}$ (bold values) is the fraction of events with a reconstructed multiplicity 
identical to the simulated multiplicity.}
\begin{center}
\footnotesize
{\renewcommand{\arraystretch}{1.15}
\begin{tabular}{c|c|c|c|c|c|c|c|c|c}
\hline \hline
Reconstructed & \multicolumn{9}{c}{Simulated multiplicity: $10^{5}$ multiple-muon 
events per multiplicity}  \\
\cline{2-10}
M & $M = 2$ & $M = 3$ & $M = 4$ & $M = 5$ & $M = 6$ & $M = 7$ & 
$M = 8$ & $M = 9$ & $M = 10$ \\ 
\hline
1 & 16593 & 15632 & 12248 & 10314 & 9351 & 8856 & 8525 & 8526 & 8411 \\
2 & \bf 58057 & 28359 & 24000 & 22080  & 17845 & 15919 & 15004 & 14550 & 14154 \\
3 & 1 & \bf 38355 & 27336 & 24906 & 22757 & 21092 & 19897 & 18999 & 18422 \\
4 & & 1 & \bf 21913 & 20577 & 20245 & 20198 & 19744 & 19024 & 18205 \\
5 & & & 1 & \bf 10654 & 12593 & 13794 & 14285 & 14428 & 14022 \\
6 & & & & & \bf 4497 & 6420 & 7532 & 8021 & 8351 \\
7 & & & & & 1 & \bf 1697 & 2686 & 3471 & 3914 \\
8 & & & & & & & \bf 502 & 999 & 1347 \\
9 & & & & & & & & \bf 131 & 319 \\
10 & & & & & & & & & \bf 33 \\
\hline
${\varepsilon_{[rec.,\, {\it M}]}}$  & 58.1\% &  66.7\% &  73.3\% & 78.2\% & 
77.9\% & 79.1\% & 79.7\% & 79.6\% & 78.8\% \\ 
\hline
${\varepsilon_{[rec.=gen.]}}$ & 58.1\% & 38.4\% & 21.9\% & 10.7\% & 4.5\% & 1.7\% & \multicolumn{3}{c}{$<$ 1\%} \\ 
\hline \hline
\end{tabular}
}
\end{center}
\end{table*}

The reconstruction algorithms must form tracks out of scintillator signals. The scintillator 
strips in alternating planes are oriented at $90^{\circ}$ to each other; scintillator hits in 
each of these two views are used to reconstruct muon tracks. The tracks of multiple 
muons passing through the detector simultaneously may overlap in one or both of these 
views. This overlap can confuse the reconstruction algorithms, resulting in 
reconstruction failures. To assess the frequency of such reconstruction failures, it is 
necessary to produce a sample of simulated muons with distributions of vertex positions 
and directions that accurately match the data. To obtain the necessary MC sample, the 
vertex positions and direction cosines of a large sample of real cosmic muon data are 
used as the seeds of simulated multiple-muon events in the detector. Once a vertex 
position and direction has been chosen for the primary muon of the event, the vertex 
positions and directions of the subsequent muons are taken from real muons that have 
directions within $5^{\circ}$ of the primary muon. To obtain broadly representative 
energy and charge distributions for these simulated muons, the energy and charge of 
each muon is taken from a CORSIKA \cite{bib:corsika} simulation, which is described in 
more detail below.

To evaluate the track reconstruction efficiency, samples of $10^{5}$ multiple-muon 
events were generated for each muon multiplicity from 2 to 10. Table~\ref{tab:trk_eff} 
shows the remaining number of reconstructed multiple-muon events in the MINOS FD 
after the preselection and analysis cuts are applied. For each generated multiplicity, 
$M$, Table~\ref{tab:trk_eff} shows the track reconstruction efficiency of the 
multiple-muon events, ${\varepsilon_{[rec.,\, {\it M}]}}$, defined as the sum of all events 
with at least two reconstructed muons divided by the number of generated 
multiple-muon events. Table~\ref{tab:trk_eff} also shows the efficiency of 
well-reconstructed multiple-muon events, ${\varepsilon_{[rec.=gen.]}}$, defined as the 
number of reconstructed multiple-muon events with the same multiplicity as the 
corresponding generated event, divided by the number of generated events. It is 
important in the analysis that multiple-muon events be identified as such, even though 
all of the muons in the event may not be reconstructed. The efficiency for identifying a 
multiple-muon event is 60\%--80\% as shown in Table~\ref{tab:trk_eff}. The efficiency for 
measuring the correct multiplicity for $M>2$ is much lower, but this is less important for 
this analysis.

The most important factor affecting the charge separation efficiency for tracks in 
multiple-muon events is the presence of large showers along the muon track, resulting 
from bremsstrahlung from radiative energy loss. The track reconstruction algorithm 
occasionally includes scintillator hits from these showers as part of the muon track, 
resulting in an otherwise straight track being reconstructed with an apparent curvature 
with high significance. The frequency of such large showers along muon tracks depends 
directly on the energy of the muon. Therefore, to obtain a sample of simulated muons 
suitable for determining the charge separation efficiency, it is vital that the energy 
distribution of the muons is correct, and in particular that the energy distribution of 
muons within each multiple-muon event is correct.

To evaluate the charge separation efficiency, the CORSIKA cosmic-ray simulation was 
used to generate the energy distribution of the muons. CORSIKA uses an initial primary 
cosmic-ray spectrum to generate particle showers in the atmosphere, and propagates 
muons from meson decay to the Earth's surface. The energy of these muons at the 
surface level is converted to energies at the detector level by considering energy loss as 
the muons traverse a distance $X$ through the Soudan rock to the detector 
\cite{bib:reichenbacher},
\begin{equation}
\label{eq:overburden}
-\frac{dE_{\mu}}{dX} = a(E_\mu) +  b(E_\mu) E_\mu,
\end{equation}
where the parameters $a$ and $b$ describe the energy lost by a muon of energy 
$E_{\mu}$ through collisional and radiative processes, respectively. Equation 
(\ref{eq:overburden}) assumes continuous energy loss and does not account for 
fluctuations~\cite{bib:lipari}. The energy loss parameters for standard rock ($a$ and 
$b$), as a function of energy, are given in~\cite{bib:reichenbacher}. The values 
considered in the analysis for these parameters are: $a = $ \unit[2.44]{MeV.cm$^2$/g} 
and $b = $ \unit[3.04 $\times$ 10$^{-6}$]{cm$^2$/g}.

A total of $1.3\times 10^{8}$ atmospheric cosmic-ray showers were generated with 
primaries in the energy range between \unit[4]{TeV} and \unit[400]{TeV}. The showers 
that resulted in multiple-muon events at the depth of the MINOS FD were kept, 
preserving the correlation between the momenta of muons within a multiple-muon event. 
These muons were then used as seeds for the GEANT4-based detector simulation, 
assigning vertex positions and directions from data events as described earlier for the 
track-efficiency MC sample. Note that the charge-ratio output from CORSIKA was not 
tuned to the MINOS data, since detector symmetries indicate there should be no 
difference in the efficiency of charge separation between a $\mu^{+}$ and a $\mu^{-}$. 
Note also that any correlation between muon energy and angle has been neglected in 
this simulation. This is acceptable, since any such correlation would have a small effect 
on the measured charge separation efficiency; in Sec.~\ref{sec:det_mcr}, a systematic 
uncertainty is determined on the charge separation efficiency that heavily dominates the 
size of any possible effect from this neglected correlation.

Defining $N^{ij}$ as the number of muons with true charge $i$ reconstructed with 
charge $j$, the charge purity, $P$, can be defined as the quotient between the number 
of well-identified muon charges ($N^{++}$ + $N^{--}$) and the total number of identified 
charges ($N^{+}$ + $N^{-}$),
\begin{equation}
\label{eq:chg_pur}
P = \frac{N^{++} + N^{--}}{N^{+} + N^{-}}.
\end{equation}

Table~\ref{tab:chg_eff} shows the number of generated, charge-separated and correctly 
identified charges as well as the charge purity obtained for several muon multiplicities.
Note that the purity obtained from the MC simulation is not calculated separately for 
positive and negative muons and that, with this definition, the purity and efficiency are 
equal. Differences in the efficiency for positive and negative muons appear in Table 
\ref{tab:chg_eff} because only the FF configuration was simulated. There is an 
asymmetry in the acceptance between tracks traveling along or against the axis of the 
detector, and a difference in overburden in those two directions. These effects cancel in 
the data using the GM.  A corrected charge ratio is obtained using the purity from Eq.
(\ref{eq:chg_pur}).

\begin{table}[tbh]
\begin{center}
\caption{\label{tab:chg_eff} The number of simulated and charge-separated muons, and 
charge purity, $P$, obtained from the MC simulation as a function of generated muon 
multiplicity, $M$.}
\footnotesize
{\renewcommand{\arraystretch}{1.1}
\begin{tabular}{c|c|c|c|c|c|c|c}
\hline \hline 
\multirow{4}{*}{$M$} & \multicolumn{2}{c|}{N$^o$ of} & 
\multicolumn{2}{c|}{N$^o$ of} & \multicolumn{2}{c|}{N$^o$ of well-} & \multirow{4}{*}{$P\,(\%)$} \\
 & \multicolumn{2}{c|}{generated\ $\mu$} & \multicolumn{2}{c|}{charge-separated\ $\mu$} & \multicolumn{2}{c|}{identified\ $\mu$} & 
 \\
\cline{2-3} \cline{4-5} \cline{6-7}
 & \multirow{2}{*}{$N_{MC}^{+}$} & \multirow{2}{*}{$N_{MC}^{-}$} & \multirow{2}{*}{$N_{MC}^{+}$} & \multirow{2}{*}{$N_{MC}^{-}$} 
& \multirow{2}{*}{$N_{MC}^{++}$} & \multirow{2}{*}{$N_{MC}^{--}$} & \multirow{2}{*}{} \\
& & & & & & &   \\
\hline 
2 & 101534 & 98466   & 2227 & 2106 & 2132 & 1979 & 94.9 $\pm$ 0.3 \\
3 & 150359 & 149641 & 2659 & 2734 & 2500 & 2523 & 93.1 $\pm$ 0.3  \\
4 & 200125 & 199875 & 3113 & 3153 & 2869 & 2889 & 91.9 $\pm$ 0.3 \\
5 & 251209 & 248791 & 3309 & 3444 & 3044 & 3160 & 91.9 $\pm$ 0.3  \\
6 & 310203 & 289797 & 3614 & 3130 & 3308 & 2781 & 90.3 $\pm$ 0.4 \\
\hline \hline
\end{tabular}
}
\end{center}
\end{table}

Previously, MINOS obtained a charge purity above 99\% for a simulated single-muon 
sample \cite{bib:adamson1}. Table~\ref{tab:chg_eff}, on the other hand, shows that 
purities for simulated multiple-muon samples are lower than those obtained for the 
single-muon events. Based on a scanning study, the lower purity of charge separation 
observed in the multimuon sample is largely due to the greater fraction of events with 
large showers. This is expected since the higher average energy of multiple-muon 
events compared to single-muon events will result in a higher rate of radiative energy 
loss.


\section{Determination of the Multiple-Muon charge ratio underground}
\label{sec:det_mcr}

As discussed in Sec. \ref{sec:data}, there is a bias in the charge ratio when it is 
calculated using only data from a single magnetic field orientation. To cancel the 
geometrical acceptance effects and alignment errors that cause this bias, data taken in 
both magnetic field configurations is combined with a GM
\cite{bib:adamson2,bib:schreiner,bib:matsuno},
\begin{equation}
R_{uncorr.} =  \Big[ \Big( \frac{N^{\mu^+}_{\rm FF}}{N^{\mu^-}_{\rm FF}}\Big) 
\times \Big( \frac{N^{\mu^+}_{\rm RF}}{N^{\mu^-}_{\rm RF}} \Big) \Big]^{1/2},
\label{eq:gm}
\end{equation}
where the $N^{\mu^\pm}_{\rm FF,RF}$ are the number of positive and negative muons 
measured in the FF and RF configurations. $R_{uncorr.}$ is the measured charge ratio, 
uncorrected for muon charge-separation efficiency. Figure \ref{fig:resol} illustrates that 
the significant bias in the charge ratio measured with a single field orientation is 
strongly suppressed in the GM.

After applying all the cuts shown in Table~\ref{tab:select}, a final sample of 312514 
muon tracks was obtained from 298291 events with $2 \leq M \leq 10$. In the FF 
sample 137392~$\mu^+$ and 115714~$\mu^-$ were selected. In the RF sample 29732 
$\mu^+$ and 29676~$\mu^-$ were selected. The mean  reconstructed momentum for 
these tracks in the detector is 48 GeV, while the typical momentum of muons that fail the 
cuts is much higher. Table~\ref{tab:multi1} shows the number of observed positive and 
negative charge-separated muons in both field configurations as a function of the muon 
multiplicity.

\begin{table}[bth]
\caption{\label{tab:multi1} Number of charge-separated muons in both field 
configurations as a function of the measured multiplicity, $M$.}
\begin{center}
\footnotesize
{\renewcommand{\arraystretch}{1.1}
\begin{tabular}{c|c|c|c|c}
\hline \hline
\multirow{2}{*}{$M$} & \multicolumn{2}{c|}{Forward Field (FF)}  & 
\multicolumn{2}{c}{Reverse Field (RF)}  \\ 
\cline{2-5}
\rule{0pt}{4ex}  & N$^{\mu^+}_{\rm FF}$ & N$^{\mu^-}_{\rm FF}$ & N$^{\mu^+}_{\rm RF}$ & N$^{\mu^-}_{\rm RF}$ \\ 
\hline
2 & 106248  & 88924  & 23282 & 22719  \\ 
3 & 20886   & 18049  & 4330  & 4594   \\ 
4 & 6501    & 5578   & 1382  & 1488   \\ 
5 & 2386    & 1972   & 457   & 534   \\ 
6 & 888     & 770    & 187   & 212   \\ 
7 & 323     & 271    & 70    & 92  \\ 
8 & 104     & 98     & 18    & 26  \\ 
9 & 42      & 42     & 5     & 9  \\ 
10 & 14     & 10     & 1     & 2  \\ 
\hline 
All  & 137392 & 115714 & 29732 & 
29676 \\ 
\hline \hline
\end{tabular}
}
\end{center}
\end{table}

Table~\ref{tab:multi2} shows the calculated muon charge ratio as a function of the muon 
multiplicity obtained from the GM of the two magnetic field orientations. The measured 
charge ratio over all multiplicities is \mbox{$R_{uncorr.}=1.091 \pm 0.005 \mathrm{\, (stat.)}$}.

To obtain the true charge ratio of the multiple-muon events reaching the MINOS FD, 
$R_{uncorr.}$ must be corrected to account for the charge-separation efficiency, 
$\varepsilon$. The details of this correction are given in Appendix \ref{append}. The 
corrected charge ratio, $R_{corr.}$, is related to the uncorrected GM, $R_{uncorr.}$, 
and the charge-separation efficiency, $\varepsilon$, from Table~\ref{tab:chg_eff}, by
\begin{equation}
\label{eq:tcr}
R_{corr.} = \frac{N^{++} + N^{+-}}{N^{--} + N^{-+}} = \frac{ R_{uncorr.} - 
(\frac{1-\varepsilon}{\varepsilon}) } {1 - R_{uncorr.} \times (\frac{1-\varepsilon}{\varepsilon})}.
\end{equation}

Table~\ref{tab:multi4} shows $R_{corr.}$ as a function of muon multiplicity, taking into 
account the fact that $\varepsilon$ depends upon the multiplicity. Over all multiplicities, 
the correction increases the charge ratio by 0.013, giving an efficiency-corrected charge 
ratio of $R_{corr.} = 1.104\pm 0.006 
{\rm \,(stat.)}$.

\begin{table}[htb]
\caption{\label{tab:multi2} Summary of the measured muon charge ratio, $R_{uncorr.}$,
as a function of measured muon multiplicity, $M$, for FF and RF data, and the GM 
combination. The errors shown on the charge ratios are only statistical.}
\small
\begin{center}
{\renewcommand{\arraystretch}{1.1}
\begin{tabular}{c|c|c|c}
\hline \hline
\multirow{2}{*}{M} & Forward field & Reverse field & Geometric mean \\ 
& (FF) & (RF) & (GM) \\
\hline 
2 &\, 1.195 $\pm$ 0.005 \, & \, 1.025 $\pm$ 0.010 \, & \, 1.107 $\pm$  0.006  \, \\ 
3 & 1.157 $\pm$  0.012 & 0.943 $\pm$ 0.020 & 1.044 $\pm$  0.012 \\ 
4 & 1.165 $\pm$  0.021 & 0.929 $\pm$ 0.035 & 1.040 $\pm$  0.022 \\ 
5 & 1.210 $\pm$  0.037 & 0.856 $\pm$ 0.055 & 1.018 $\pm$  0.036 \\ 
6 & 1.153 $\pm$  0.057 & 0.882 $\pm$ 0.088 & 1.009 $\pm$  0.056 \\ 
7 & 1.192 $\pm$  0.098 & 0.761 $\pm$ 0.121 & 0.952 $\pm$  0.085 \\ 
8 & 1.061 $\pm$  0.149 & 0.692 $\pm$ 0.212 & 0.857 $\pm$  0.145 \\ 
9 & 1.000 $\pm$  0.218 & 0.556 $\pm$ 0.310 & 0.745 $\pm$  0.223 \\ 
10 & 1.400 $\pm$  0.580 & 0.500 $\pm$  0.612 & 0.837 $\pm$  0.541 \\ 
\hline
All & 1.187 $\pm$  0.005 & 1.002 $\pm$  0.008 & 1.091 $\pm$  0.005 \\ 
\hline \hline
\end{tabular}
}
\end{center}
\end{table}

\begin{table}[htb]
\small
\caption{\label{tab:multi4} Efficiency-corrected charge ratios as a function of measured 
muon multiplicity, $M$.}
{\renewcommand{\arraystretch}{1.2}
\begin{tabular}{c|c|c|c}
\hline \hline
\multirow{2}{*}{$M$}  & Meas. charge ~ & Charge & Corrected charge\\ 
  & ratio ($R_{uncorr.}$) &  efficiency (\%) & ratio ($R_{corr.}$) \\ \hline
2 & \, 1.107 $\pm$ 0.006 \, & \, 94.9 $\pm$ 0.3 \, & \, 1.119 $\pm$  0.007 \, \\ 
3 & 1.044 $\pm$  0.012  & 93.1 $\pm$  0.3  & 1.052 $\pm$  0.014  \\ 
4 & 1.040 $\pm$  0.022  & 91.9 $\pm$  0.3  & 1.048 $\pm$  0.026  \\ 
5 & 1.018 $\pm$  0.036  & 91.9 $\pm$  0.3  & 1.021 $\pm$  0.043  \\ 
6 & 0.974 $\pm$  0.044  & 90.3 $\pm$  0.4  & 0.968 $\pm$  0.054  \\ 
\hline
All & 1.091 $\pm$ 0.005  & 94.4 $\pm$  0.3  & 1.104 $\pm$  0.006  \\ 
\hline \hline
\end{tabular}
}
\end{table}

Two sources of systematic error are considered: first, a contribution from possible 
failure to fully cancel effects of magnetic field and alignment errors by reversing the 
magnetic field (bias); second, a contribution from not fully accounting for the 
charge-separation failures that tend to give a random charge determination 
(randomization) \cite{bib:schreiner}.

The systematic error on bias can be evaluated by comparing the ratio 
$N^{\mu^{+}}_{\rm FF}/N^{\mu^{-}}_{\rm RF}$ to the ratio 
$N^{\mu^{+}}_{\rm RF}/N^{\mu^{-}}_{\rm FF}$ which, in the case of no bias, should be 
identical. This comparison accounts for all biases whatever the source, and includes
focusing effects, errors in the magnetic field maps, and possible curvatures in the 
coordinate system. This systematic error was determined for the MINOS FD 
single-muon charge-ratio analysis~\cite{bib:adamson2} to be $\pm 0.009$, and this 
value of the uncertainty also applies to this multiple-muon analysis.

To calculate the systematic uncertainty on the rate of charge randomization, the error on 
the measured charge misreconstruction rate in the MC simulation is estimated. These 
charge reconstruction failures are dominated by events with large radiative energy loss, 
which has a significantly higher rate for muon energies above \unit[1]{TeV}.  An 
inaccurate muon energy distribution being modeled by CORSIKA would be a source of
systematic error. This is examined by comparing some features of our MC with two other 
MC simulations: a different version of CORSIKA, and an earlier program developed for 
the Soudan 2 experiment \cite{bib:kasahara,bib:kasahara2}, which studied multiple-muon 
events at a location near the MINOS FD. No differences were noticed in the calculations 
of energy loss, multiplicity, and other features of multiple-muon events underground.  As 
another check, the rate of reconstructed showers was compared in the data, in the MC 
simulation, and in charge-misidentified MC events. There was a negligible rate of 
charge-misidentified events with no showers.  The mean number of showers in events 
with at least one shower in these three samples was 1.70, 2.14 and 2.68 respectively.  
There is thus some evidence that the Monte Carlo simulation is overestimating the 
number of high-energy muon events, and hence the correction to the charge ratio. The 
ratio 2.14/1.70 = 1.26 is taken as evidence that there are 26\% more showers in the MC
than there should be. This value is conservatively increased by half, and 39\% 
systematic error is used as a possible overcorrection. Half of 39\%, or 20\%, is then 
taken as the systematic error on a possible undercorrection. Since the size of the 
correction is 0.013, this leads to a systematic error on $R_{corr.}$ from randomization 
of $^{+0.003}_{-0.005}$. When combined in quadrature with the systematic error from 
bias, the total systematic error is  $^{+0.009}_{-0.010}$.

Thus the efficiency-corrected multiple-muon charge ratio at a depth of 2070 mwe is 
determined to be 
$R_{corr.} = 1.104\pm 0.006 {\rm\,(stat.)}^{+0.009}_{-0.010}{\rm\,(syst.)}$.
This measurement agrees within uncertainties with the recent OPERA measurement 
of $1.098 \pm 0.023{\rm\, (stat.)}^{+0.015}_{-0.013}{\rm\, (syst.)}$ (2014) 
\cite{bib:opera2} and has a much smaller uncertainty.


\section{Summary}
\label{sec:summary}
A measurement of the multiple-muon charge ratio, $R_\mu = N_{\mu^+}/N_{\mu^-}$, 
has been performed using the full MINOS FD atmospheric data set.  For 
multiple-muon events the measured charge ratio is 
$R_{uncorr.} = 1.091 \pm 0.005 {\rm\, (stat.)}$ before correcting for charge 
misidentification. The efficiency-corrected charge ratio is
\begin{equation}
\label{eq:ccr}
R_{corr.} = 1.104 \pm 0.006 {\rm\, (stat.)} ^{+0.009}_{-0.010} {\rm\, (syst.)}.
\end{equation}

The calculated underground multiple-muon charge ratio [Eq. \ref{eq:ccr})] is lower than 
the single-muon charge ratio measurements obtained by several experiments in the 
past \cite{bib:adamson1,bib:achard,bib:haino,bib:zimmermann,bib:khachatryan,
bib:hebbeker,bib:adamson2,bib:opera1}. This result gives support to hypotheses about 
the decrease of the charge ratio for multiple-muon events discussed in Sec.
\ref{sec:intro}, providing a better understanding of the mechanism of multiple-muon 
production in the atmosphere. Although the measured ratio in principle depends on the 
depth, shape of the overburden, area, and MDM of an underground detector, the result 
is consistent with the last OPERA multiple-muon charge ratio measurement
\cite{bib:opera2}.


\section{ACKNOWLEDGMENTS}

This work was supported by the U.S. DOE, the U.K. STFC, the U.S. NSF, the State and 
University of Minnesota, the University of Athens, Greece, and Brazil's FAPESP, CAPES 
and CNPq. We are grateful to the Texas Advanced Computer Center, the Minnesota 
Department of Natural Resources, the crew of Soudan Underground Laboratory, and 
the Fermilab personnel for their contributions to this effort.

\vspace*{25pt}

\appendix

\section{Purity-corrected multiple-muon charge ratio}
\label{append}

As defined in Sec. \ref{sec:mc}, $N^{ij}$ is the number of muons with true charge $i$ 
reconstructed with charge $j$. Assuming that the charge efficiency, $\varepsilon$, is the 
same for both positive and negative muons in both MC and data, we have
\begin{equation}
\varepsilon = \frac{N^{++}}{N^{++} + N^{+-}} = \frac{N^{--}}{N^{--} + 
N^{-+}}, \label{eq:eff}
\end{equation}
\begin{equation}
N^{+-} = N^{++} \times \Big( \frac{1-\varepsilon}{\varepsilon} \Big),
\label{eq:ndnp}
\end{equation}
\begin{equation}
N^{-+} = N^{--} \times \Big( \frac{1-\varepsilon}{\varepsilon} \Big). 
\label{eq:ndpn}
\end{equation}

Combining Eqs. (\ref{eq:eff}), (\ref{eq:ndnp}) and (\ref{eq:ndpn}) we can express the 
measured charge ratio as
\begin{equation}
R_{uncorr.} = \frac{N^{++} + N^{-+}}{N^{--} + N^{+-}} = \frac{ 
\frac{N^{++}}{N^{--}} + (\frac{1-\varepsilon}{\varepsilon}) } 
{1 + \frac{N^{++}}{N^{--}} \times (\frac{1-\varepsilon}{\varepsilon})}.
\end{equation}
Reordering the terms,
\begin{equation}
\frac{N^{++}}{N^{--}}  = \frac{ R_{uncorr.} - (\frac{1-\varepsilon}
{\varepsilon}) } {1 - R_{uncorr.} \times (\frac{1-\varepsilon}{\varepsilon})}.
\label{eq:tc}
\end{equation}
Furthermore, the true charge ratio is
\begin{eqnarray}
R_{corr.} & = & \frac{N^{++} + N^{+-}}{N^{--} + N^{-+}} \nonumber \\
& = & \frac{N^{++} + N^{++} \times (\frac{1-\varepsilon}{\varepsilon}) } 
{N^{--} + N^{--} \times (\frac{1-\varepsilon}{\varepsilon})} = \frac{N^{++}}
{N^{--}}. 
\label{eq:tcr1}
\end{eqnarray}

Combining Eqs. (\ref{eq:tc}) and (\ref{eq:tcr1}), the corrected  charge ratio is given by
\begin{eqnarray}
R_{corr.} & = & \frac{ R_{uncorr.} - (\frac{1-\varepsilon}{\varepsilon}) } 
{1 - R_{uncorr.} \times (\frac{1-\varepsilon}{\varepsilon})}.
\label{eq:tcr2}
\end{eqnarray}
 
The associated error, $\delta  R_{corr.}$, is obtained by propagating the errors on 
$R_{uncorr.}$ and $\varepsilon$ through Eq. (\ref{eq:tcr2}):
\begin{equation}
\textstyle
\delta R_{corr.} =  \frac{  \sqrt{(1-2\varepsilon)^2 \times 
(\delta R_{uncorr.})^2 + (1-R^2_{uncorr.}) \times (\delta \varepsilon)^2} }
{ \big[\varepsilon - R_{uncorr.} \times (1-\varepsilon) \big]^2}
\label{eq:etcr2}.
\end{equation}
 Using $\varepsilon = 0.944$ and $R_{uncorr.} = 1.091$, we obtain
$\delta R_{corr.} = 0.006{\rm\,(stat.)}$.

\nocite{*}

\bibliography{references}

\end{document}

%% file: ab.tex
\newcommand{\Berkeley}{Lawrence Berkeley National Laboratory, Berkeley, California, 94720 USA}
\newcommand{\Cambridge}{Cavendish Laboratory, University of Cambridge, Madingley Road, Cambridge CB3 0HE, United Kingdom}
\newcommand{\Cincinnati}{Department of Physics, University of Cincinnati, Cincinnati, Ohio 45221, USA}
\newcommand{\FNAL}{Fermi National Accelerator Laboratory, Batavia, Illinois 60510, USA}
\newcommand{\RAL}{Rutherford Appleton Laboratory, Science and Technology
 Facilities Council, Didcot, OX11 0QX, United Kingdom}
\newcommand{\UCL}{Department of Physics and Astronomy, University College London, Gower Street, London WC1E 6BT, United Kingdom}
\newcommand{\Caltech}{Lauritsen Laboratory, California Institute of Technology, Pasadena, California 91125, USA}
\newcommand{\Alabama}{Department of Physics and Astronomy, University of Alabama, Tuscaloosa, Alabama 35487, USA}
\newcommand{\ANL}{Argonne National Laboratory, Argonne, Illinois 60439, USA}
\newcommand{\Athens}{Department of Physics, University of Athens, GR-15771 Athens, Greece}
\newcommand{\NTUAthens}{Department of Physics, National Tech. University of Athens, GR-15780 Athens, Greece}
\newcommand{\Benedictine}{Physics Department, Benedictine University, Lisle, Illinois 60532, USA}
\newcommand{\BNL}{Brookhaven National Laboratory, Upton, New York 11973, USA}
\newcommand{\CdF}{APC -- Universit\'{e} Paris 7 Denis Diderot, 10, rue Alice Domon et L\'{e}onie Duquet, F-75205 Paris Cedex 13, France}
\newcommand{\Cleveland}{Cleveland Clinic, Cleveland, Ohio 44195, USA}
\newcommand{\Delhi}{Department of Physics \& Astrophysics, University of Delhi, Delhi 110007, India}
\newcommand{\GEHealth}{GE Healthcare, Florence South Carolina 29501, USA}
\newcommand{\Harvard}{Department of Physics, Harvard University, Cambridge, Massachusetts 02138, USA}
\newcommand{\HolyCross}{Holy Cross College, Notre Dame, Indiana 46556, USA}
\newcommand{\Houston}{Department of Physics, University of Houston, Houston, Texas 77204, USA}
\newcommand{\IIT}{Department of Physics, Illinois Institute of Technology, Chicago, Illinois 60616, USA}
\newcommand{\Iowa}{Department of Physics and Astronomy, Iowa State University, Ames, Iowa 50011 USA}
\newcommand{\Indiana}{Indiana University, Bloomington, Indiana 47405, USA}
\newcommand{\ITEP}{High Energy Experimental Physics Department, ITEP, B. Cheremushkinskaya, 25, 117218 Moscow, Russia}
\newcommand{\JMU}{Physics Department, James Madison University, Harrisonburg, Virginia 22807, USA}
\newcommand{\LASL}{Nuclear Nonproliferation Division, Threat Reduction Directorate, Los Alamos National Laboratory, Los Alamos, New Mexico 87545, USA}
\newcommand{\Lebedev}{Nuclear Physics Department, Lebedev Physical Institute, Leninsky Prospect 53, 119991 Moscow, Russia}
\newcommand{\LLL}{Lawrence Livermore National Laboratory, Livermore, California 94550, USA}
\newcommand{\LosAlamos}{Los Alamos National Laboratory, Los Alamos, New Mexico 87545, USA}
\newcommand{\Manchester}{School of Physics and Astronomy, University of Manchester, Oxford Road, Manchester M13 9PL, United Kingdom}
\newcommand{\MIT}{Lincoln Laboratory, Massachusetts Institute of Technology, Lexington, Massachusetts 02420, USA}
\newcommand{\Minnesota}{University of Minnesota, Minneapolis, Minnesota 55455, USA}
\newcommand{\Crookston}{Math, Science and Technology Department, University of Minnesota -- Crookston, Crookston, Minnesota 56716, USA}
\newcommand{\Duluth}{Department of Physics, University of Minnesota Duluth, Duluth, Minnesota 55812, USA}
\newcommand{\Ohio}{Center for Cosmology and Astro Particle Physics, Ohio State University, Columbus, Ohio 43210 USA}
\newcommand{\Otterbein}{Otterbein College, Westerville, Ohio 43081, USA}
\newcommand{\Oxford}{Subdepartment of Particle Physics, University of Oxford, Oxford OX1 3RH, United Kingdom}
\newcommand{\PennState}{Department of Physics, Pennsylvania State University, State College, Pennsylvania 16802, USA}
\newcommand{\PennU}{Department of Physics and Astronomy, University of Pennsylvania, Philadelphia, Pennsylvania 19104, USA}
\newcommand{\Pittsburgh}{Department of Physics and Astronomy, University of Pittsburgh, Pittsburgh, Pennsylvania 15260, USA}
\newcommand{\IHEP}{Institute for High Energy Physics, Protvino, Moscow Region RU-140284, Russia}
\newcommand{\Rochester}{Department of Physics and Astronomy, University of Rochester, New York 14627 USA}
\newcommand{\RoyalH}{Physics Department, Royal Holloway, University of London, Egham, Surrey, TW20 0EX, United Kingdom}
\newcommand{\Carolina}{Department of Physics and Astronomy, University of South Carolina, Columbia, South Carolina 29208, USA}
\newcommand{\SDakota}{South Dakota School of Mines and Technology, Rapid City, South Dakota 57701, USA}
\newcommand{\SLAC}{Stanford Linear Accelerator Center, Stanford, California 94309, USA}
\newcommand{\Stanford}{Department of Physics, Stanford University, Stanford, California 94305, USA}
\newcommand{\StJohnFisher}{Physics Department, St. John Fisher College, Rochester, New York 14618 USA}
\newcommand{\Sussex}{Department of Physics and Astronomy, University of Sussex, Falmer, Brighton BN1 9QH, United Kingdom}
\newcommand{\TexasAM}{Physics Department, Texas A\&M University, College Station, Texas 77843, USA}
\newcommand{\Texas}{Department of Physics, University of Texas at Austin, 1 University Station C1600, Austin, Texas 78712, USA}
\newcommand{\TechX}{Tech-X Corporation, Boulder, Colorado 80303, USA}
\newcommand{\Tufts}{Physics Department, Tufts University, Medford, Massachusetts 02155, USA}
\newcommand{\UNICAMP}{Universidade Estadual de Campinas, IFGW-UNICAMP, CP 6165, 13083-970, Campinas, SP, Brazil}
\newcommand{\UFG}{Instituto de F\'{i}sica, 
Universidade Federal de Goi\'{a}s, 74690-900, Goi\^{a}nia, GO, Brazil}
\newcommand{\USP}{Instituto de F\'{i}sica, Universidade de S\~{a}o Paulo,  CP 66318, 05315-970, S\~{a}o Paulo, SP, Brazil}
\newcommand{\Warsaw}{Department of Physics, University of Warsaw, Pasteura 5, PL-02-093 Warsaw, Poland}
\newcommand{\Washington}{Physics Department, Western Washington University, Bellingham, Washington 98225, USA}
\newcommand{\WandM}{Department of Physics, College of William \& Mary, Williamsburg, Virginia 23187, USA}
\newcommand{\Wisconsin}{Physics Department, University of Wisconsin, Madison, Wisconsin 53706, USA}
\newcommand{\deceased}{Deceased.}

\affiliation{\ANL}
\affiliation{\BNL}
\affiliation{\Caltech}
\affiliation{\Cambridge}
\affiliation{\UNICAMP}
\affiliation{\Cincinnati}
\affiliation{\FNAL}
\affiliation{\UFG}
\affiliation{\Harvard}
\affiliation{\HolyCross}
\affiliation{\Houston}
\affiliation{\IIT}
\affiliation{\Indiana}
\affiliation{\Iowa}
\affiliation{\UCL}
\affiliation{\Manchester}
\affiliation{\Minnesota}
\affiliation{\Duluth}
\affiliation{\Otterbein}
\affiliation{\Oxford}
\affiliation{\Pittsburgh}
\affiliation{\RAL}
\affiliation{\USP}
\affiliation{\Carolina}
\affiliation{\Stanford}
\affiliation{\Sussex}
\affiliation{\TexasAM}
\affiliation{\Texas}
\affiliation{\Tufts}
\affiliation{\Warsaw}
\affiliation{\WandM}

\author{P.~Adamson}
\affiliation{\FNAL}


\author{I.~Anghel}
\affiliation{\Iowa}
\affiliation{\ANL}



\author{A.~Aurisano}
\affiliation{\Cincinnati}









\author{G.~Barr}
\affiliation{\Oxford}









\author{M.~Bishai}
\affiliation{\BNL}

\author{A.~Blake}
\altaffiliation[Now at\ ]{Lancaster University, Lancaster, LA1 4YB, UK.}
\affiliation{\Cambridge}

\author{G.~J.~Bock}
\affiliation{\FNAL}


\author{D.~Bogert}
\affiliation{\FNAL}




\author{S.~V.~Cao}
\affiliation{\Texas}

\author{T.~J.~Carroll}
\affiliation{\Texas}

% ...

\collaboration{The MINOS Collaboration}
\noaffiliation